# Presenting a Method for Improving Echo Hiding

Haniyeh Rafiee[1]        Mohammad Fakhredanesh[*2]

**Abstract**: In this article, one of the most important methods of steganography on VoIP called echo hiding is improved. This method has advantages in maintaining the statistical and perceptual characteristics of audio signals as well as security against the sensitivity of human audio system (HAS). However, it has lots of errors in detecting coded and hidden messages, which is detectable using existing steganalysis methods. The percentage of extracting messages in these improved methods of echo hiding is high, but they lower the security of the method. In this article, a method is presented to improve the method of extracting echo hiding, and enhance its security through a combined method based on spread spectrum. To improve the extraction, a wrong hypothesis is corrected and substituted. To improve security using a pseudo-random key generation algorithm, spread spectrum and echo hiding methods are used randomly. To evaluate the proposed extraction, numerous extraction tests are carried out in the normal state and in the event of attacks. A steganalyser has also been used to assess security improvements. The results gained through different experiments on the security of steganography indicate a 3-percent increase in steganalysis errors. The proposed extraction method was modified based on the main method and resulted in more than 10% improvement.

**Keywords**: Steganography, echo hiding, spread spectrum, security, Cepstral coefficients, self-symmetric

## 1. Introduction

Steganography is a kind of secret communication method in which the message is hidden in a cover file and detecting the messages in digital media is almost impossible. The file can be an image, audio, video, and text file, or Internet telephone call. The hiding method for phone calls is different.

VoIP phone is one of the most favored Internet services substituted for public telephone network [1]. Due to the abundant use in the Internet, voice calls are suitable signal cover for steganography [2]. Steganography methods used for VoIP are divided in two major groups. In the first method, the sound is used as the hidden data carrier, while in the second method the protocol is introduced as the carrier. Some of the methods belong to both groups. Some methods in one group may be improved based on a specific technique. In this article, we particularly focus on improving one of the stenographical methods on VoIP which uses sound as the carrier. This method is called echo hiding introduced by Gruhel et al. [3]. In this method, secret data are embedded by introducing a short echo to the host signal. Erfani and Siyahpoosh [4] also introduced time-spread echo hiding method and offered ways to improve it. Chen et al [5] also introduced a safe method on the basis of echo hiding. Kim et al. [6] attempted to improve echo hiding by adding two types of delay from the back and front to the kernel. Then Delfrozi and Pouyan offered a new way to increase the capacity of this method [7].

Adding negative and positive echoes from the back to the kernel is also another way to improve echo hiding [8]. One of the most important improvements is called mirrored echo hiding [9].

In this research, the extraction and security of echo hiding is improved. Extraction is improved by correcting the autocorrelation between cepstral coefficients and also by rejecting the hypothesis proposed on the basis of self-symmetry of the signal. The extracting method can also be improved through reviewing the relationship between equations. Security is also improved by combining the proposed algorithm using the spread spectrum approach. Combation is based on using a pseudo-random key generation algorithm. This algorithm has a prime key and many subkeys. Subkeys are aslo generated by the prime key.

The present article is composed of six main parts. After the introduction and literature review parts, the third part explains how to embed the echo hiding method and extract it. The fourth part proceeds with improving the rate of message recovery in echo hiding method and ends with suggesting methods to improve it. In the fifth part, simulations and the results gained through simulations are discussed. A general summary of the article is also presented in the last part.

## 2. Literature Review

The first method of VoIP steganography which has chosen sound as the carrier was proposed in 2003 [10]. In this method, side information about the sound wave made with G.711 codec was sent from the receiver by steganography LSB (least significant byte). Later, other methods were proposed on the basis of steganography in LSB. Wang and Wu [11] were among those who suggested a method of adaptive steganography with least significant byte (LSB) in which the LSBs suitable for carrying data are estimated and steganography only takes place in those bytes. In 2009, other adaptive methods on LSBs were proposed [12]. In 2011, another adaptive steganography was offered based on the smoothness of block tall [13]. Further research was carried out to upgrade adaptive steganography in least significant byte (LSB), which then becomes an embeded steganography system called APMS. The improved versions are called APMS1 and APMS2 [14, 15].

Using VoIP specific protocols as the carrier of secret data was first proposed in 2006, [16] and then RTP and RTCP, other steganography methods for signaling protocol SIP and their sequels, were proposed [17,1]. Another method called

LACK was also suggested within the field [2]. In 2009, a secret channel was proposed on the basis of jitter field which was related to header protocol RTCP [18], and then another steganography method was offered on the basis of RTP [19]. In the same year, a new steganography method called interference channel was offered [20]. Finally, in 2012 one of the most important methods in this field was proposed: TramSteg [21]. Another complex method using sound as data carrier is the basic principle of QIM. This principle is based on using redundancy at the stage of quantization, and embedding confidential messages by altering the quantized values. This was first applied to hide photographic intelligence to reach conversation steganography with low byte rate [22]. Phase coding method is another way which applies hearing system for steganography. The human ear hardly understands the absolute coefficients of the phase, but it is sensitive to relative values. In this regard, steganography using phase coding method has utilized this characteristic of human hearing system.

Therefore, it is possible to change the phase of the first part of the audio file with a reference phase used for steganography and then adjust the phase of other parts so that the relative phase between different parts remains untouched [23]. Cox et al. proposed a very strong method called spread spectrum which plays the bytes of secret messages using the key in the audio file [24].

Recently, the scope of steganography on VoIP has also entered new technologies. For example, Yijing et al. [25] worked on steganography in VoIP communication using smart grids. This method gained good results against statistical steganalysis. In this approach, pulse code modulation (PCM) technique is used for embedding method. Similiraly, Zhanzhan Gao et al. [26] presented a method for small and large payloads of packet VoIP in their recent study. This method uses combining Hamming code into parity check matrix for small payloads and combining parity check matrix of syndrome trellis codes with several referential columns for large payloads. In another recent work, Li et al. [27] presented a steganographic method using the excitation pulse positions of high bit rate speech codec of ITU-T G.723.1.

As mentioned before, echo hiding and its improved versions are amongst other important methods of steganography. In this method, the message is embedded in the host signal using the echo and tow delays [3]. Since the message is not extracted in this method, several modulations have been carried out to improve it, which will be discussed further on.

## 3. Introducing echo hiding and its improvements

Echo hiding is one of the cryptographic methods on audio media, such as music or recorded audio. The feasibility of this method in VoIP protocol is approved. This method can maintain the sound quality as well as offering a bandwidth of 20 byte/second [28]. As was discussed before, the method has significant advantages in maintaining statistical and perceptual features of the sound, and it is sensitive to human audio system (HAS). Different improvements have been made on this method to make it more applicable. In section 3.1, we will discuss how the method is embedded and how it is improved later. In section 3.2, we will talk about extraction.

### 3.1. Embedding echo hiding and its improvment

In this method, the bytes of the message are embedded in the host signal using different delays and defining one or two echo cores. If both echoes are under the threshold of human hearing, steganogarphy will not produce a significant noise. For embedding with this method, each part of host signal should go through existing delays and the mentioned byte should be 0 or 1 [3].

The core of the echo according to this method should be as follows [3]:

$$h(n) = \delta(n) + \alpha.\delta(n-d) \tag{1}$$

Where n is the number of samples, $\delta$ the beat, and $\alpha$ and d are echo and delay coefficients, respectively. Therefore, echoed signal is [3]:

$$s(n) = x(n) * h(n) = x(n) + \alpha.x(n-d), \tag{2}$$

* indicates convolution, and x (n) and s (n) are the host signal and its echoed version, respectively.

Ooh et al. offered an improvised version of echo hiding which was composed of positive and negative pulses with short differences. The formula for the core in this method is as follows [8]:

$$h(n) = \delta(n) + \alpha_{PB}.\delta(n-d_{PB}) - \alpha_{NB}.\delta(n-d_{NB}), \tag{3}$$

In the core, $a_{PB}$ and $a_{NB}$ are the coefficients of positive echo backward and negative echo backward, respectively. $d_{PB}$ And $a_{NB}$ show positive delay backward and negative delay backward, respectively, and the space between these two should be from 1 to 5.

Kim et al. offered a new core of echo hiding. In this method, two pulses were added forward and backward. The formula of the core is as follows [7]:

$$h(n) = \delta(n) + \alpha.\delta(n-d) + \alpha.\delta(n+d), \tag{4}$$

The first part is the one added to the signal of the echo range in the backward pulse and the second part is the one added to the echo range in the forward pulse.

Next improvement on echo hiding is to apply reflection. In this method, the echo range in positive and negative pulses is added to the backward and forward signals. In this method, the formula is explained in [9]:

$$h(n) = \delta(n) + \alpha_{PB}.\delta(n-d_{PB}) + \alpha_{PB}.\delta(n+d_{PB}) \\ -\alpha_{NB}.\delta(n-d_{NB}) - \alpha_{NB}.\delta(n+d_{NB}). \tag{5}$$

The last improvement on the echo hiding method is time-spread method. In this method, the coefficient of the echo is highly close to zero [4]:

$$h(n) = \delta(n) + \alpha.p(n-d), \tag{6}$$

In the formula, p is a pseudorandom noisy function. Regarding the improvement applied, the core formula is as follows [4]:

$$h(n) = \delta(n) + \alpha . \sum_{i=1}^{N} p(i).x(n-d-i). \tag{7}$$

All improvements mentioned above were used for robustness and increasing the rate of improvement in echo hiding method. The aforementioned methods have a higher rate of recovery than the echo hiding method, but they mostly lack suitable measures to increase the security.

### 3.2. Echo hiding extraction
In order to reveal the embedded echo, the peaks of two probable locations in the cepstrum zone are compared with each other. Mixed cepstrum in d0 and d1 will have an instant peak indicating echo. Mixed cepstrum for this signal is as follows [3]:

$$
\begin{aligned}
S(n) &= FT^{-1}\left[\ln\left(FT\left[s(n)\right]\right)\right] \\
&= FT^{-1}\left[\ln\left(FT[x(n)]\right)\right] + FT^{-1}\left[\ln\left(FT[h(n)]\right)\right] \\
&= FT^{-1}\left[\ln\left(X\left(e^{jw}\right)\right)\right] + FT^{-1}\left[\ln\left(H\left(e^{jw}\right)\right)\right] \\
&= c_x(n) + c_h(n).
\end{aligned}
\tag{8}
$$

Were

$$
\begin{aligned}
c_x(n) &= FT^{-1}\left[\ln\left(X\left(e^{j\Omega}\right)\right)\right], \\
c_h(n) &= FT^{-1}\left[\ln\left(H\left(e^{j\Omega}\right)\right)\right], \\
H\left(e^{jw}\right) &= FT\left[h(n)\right] = 1 + \alpha e^{-jwd}.
\end{aligned}
$$

On the other hand

$$\ln(1+\alpha) = \alpha - \frac{\alpha^2}{2} + \frac{\alpha^3}{3} - \dots \quad , when |\alpha| < 1,$$

Therefore [3]:

$$
\begin{aligned}
\ln\left(H\left(e^{jw}\right)\right) &= \alpha.e^{-jwd} - \frac{\alpha^2}{2}.e^{-j2wd} + \frac{\alpha^3}{3}.e^{-j3wd} - \dots, \\
c_h(n) &= FT^{-1}\left[\alpha.e^{-jwd} - \frac{\alpha^2}{2}.e^{-j2wd} + \frac{\alpha^3}{3}.e^{-j3wd} - \dots\right] \\
&= \alpha.\delta(n-d) - \frac{\alpha^2}{2}.\delta(n-2d) + \frac{\alpha^3}{3}.\delta(n-3d) - \dots.
\end{aligned}
\tag{9}
$$

The result shows that a series of pulses with domains that exponentially disappear have alternately appeared. But the first hit was prominent in n=d and has the $\alpha$ domain. It is also observed that the echo should be carefully chosen to be impalpable and have a very high probability of detection. However, the results of cepstrum is not convincing in practice. The peak of cepstrum is smaller than the host signal and it may cover adjacent peaks. Therefore, autocorrelation of the cepstrum is calculated for the recovery of the messages [3].

$$R_{xx}[n] = FT^{-1}\left[\ln\left(FT\left[s(n)\right]\right)^2\right]. \tag{10}$$

Here, the first peak is significantly strengthened. In the end, the embedded byte is recognized through the comparison made between $s(d_1)$ and $s(d_0)$. If $s(d_1) > s(d_0)$, most likely the hidden byte is one, and if $s(d_1) < s(d_0)$, most likely the hidden byte is zero. The probability of equality between $s(d_1)$ and $s(d_0)$ is very insignificant; however, if it occurs, we consider hidden byte as 1 [3]. For function x[n], correlation is defined as follows [3]:

$$R_{xx}[n] = \sum_{m=-\infty}^{+\infty} x[n+m]x[m]. \tag{11}$$

If we have k=n+m, changing the variable in the equation will lead to:

$$R_{xx}[n] = \sum x[k]x[k-n], \tag{12}$$

If we factor out a minus in the second equation, then we will have:

$$R_{xx}[n] = \sum x[k]x[-(n-k)], \tag{13}$$

The equation of the convolution is defined as:

$$R_{xx}[n] = \sum x[k]x[-(n-k)], \tag{14}$$

There are similarities between the definition of convolution equation (14) and the improved correlation equation (13), and the only difference is a minus in the second. The autocorrelation can be shown as [3]:

$$R_{xx} = x[n] * x[-n]. \tag{15}$$

Supposing that x[n] signal is self-symmetric, we will have: x[n] = x [-n]; therefore, the equation of autocorrelation can be written as [3]:

$$R_{xx} = x[n] * x[n]. \tag{16}$$

If the above equation is brought to frequency domain using Fourier transform, then we have [3]:

$$S_{xx}\left(e^{j\Omega}\right) = X\left(e^{j\Omega}\right)^2, \tag{17}$$

Where $S_{xx}$ is equal to Fourier transform of $R_{xx}$ and $X^2$ is equal to Fourier transform of x[n]².
Therefore, in order to calculate cepstrum, first the signal should be taken to frequency domain using Fourier transform and then another Fourier inverse transform is needed for the signal time domain. Figure 1 shows the stages mentioned in (10) in order to calculate autocorrelation coefficient:

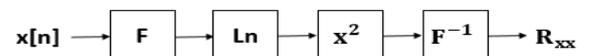

Fig. 1. System displaying autocorrelation coefficient of cepstral.
[3]

### 4. Introducing Spread Spectrum
In this method, covering signal is divided into frames with length of N; it is the same length as the key of steganography and a byte is hidden in each frame. The key of steganogarphy is pseudo-random vector formed of plus and minus 1. In each frame, steganography is as follows [24]:

$$y = x + Abs \tag{18}$$

In this equation, $x_{N \times 1}$ is the vector of covering signal, $y_{N \times 1}$ is the vector of output Stego signal, A is the embedding strength, and b is the byte of information which is equal to 1 for 1 byte and -1 for 0 byte. $s_{N \times 1}$ is the key of steganography whose elements are chosen randomly from {-1, +1}.

For revealing mass cryptographic signals, maximum likelihood estimation is used [32]:

$$\hat{b} = sgn(z_{GML}) = sgn(y^T s) \tag{19}$$

In order to decrease errors (errors happening in the channel), an improved method of mass spread spectrum was proposed in [30]. In this method, steganogarphy is done as follows:

$$y = x + Abs - kss^T x \tag{20}$$

For a better evaluation of this method and because the value of $s^T x$ is scalar, it is rewritten as:

$$\begin{aligned} y &= x + Abs - ks^T xs \\ &= x + \left(Ab - ks^T x\right)s \\ &= x + \alpha s \end{aligned} \tag{21}$$

Where k is calculated to minimize the probability of error and ɑ in each frame is different and calculated on the basis of cover vector, key, hidden byte, and the power of steganography. Therefore, we can assume that the improved spread spectrum is a mass method with a non-static steganogarphy domain.

This method is popular for noisy channels as well as the Internet telephone channels such as Skype. The feasibility of spread spectrum in VoIP protocol is evaluated, and we can reach a very strong protocol; however, the bandwidth is not high (less than 20 byte/s) [28].

## 5. Proposed Method for Echo Hiding
In this section, a method for improving the existing Echo hiding method is proposed. This section is divided in two separate ones: the first is about extraction, and the second part is about how to improve the security of echo hiding.

### 5.1. Extraction Method Proposed for Echo Hiding
Equations (15) occurs when autocorrelation is used for detection. Since the signal is supposed to be self-symmetric, equation (15) is changed into equation (16), where x [-n] = x[n]. But as only a few functions are self-symmetric, we cannot consider each type of signal as self-symmetric. On the other hand, despite this assumption we see that the rate of message recovery in the extraction of echo hiding is very low. For a better message recovery, we have to increase the value of the ɑ component in equation (2). The more the component increases, the higher the rate of recovery becomes. However, it reduces the security of this method. In this article, a pattern is proposed, which will increase the rate of message recovery while maintaining the security. Therefore, we will discuss autocorrelation while ignoring self-symmetry.

In the pattern of extraction improvement, there is x[n] ≠x [-n], and the signal samples x [-n] or x[n] are real. Therefore, by applying complex conjugate, we have the following:

$$x[-n] = x[-n]^* . \tag{22}$$

Now if we rewrite the equation of autocorrelation, we will come up with the following:

$$R_{xx} = x[n] * x[-n]^* , \tag{23}$$

The ∗ shows convolutional action, and $^*$ shows complex conjugate. If we take the equation to frequency domain by applying Fourier transform, we will have the following:

$$S_{xx}\left(e^{j\Omega}\right) = \left|X\left(e^{j\Omega}\right)\right|^2 . \tag{24}$$

Therefore, for message recovery instead of formula (10), we have the following:

$$R_{xx}[n] = FT^{-1}\left[\left|\ln\left(FT\left[s(n)\right]\right)\right|^2\right] . \tag{25}$$

Figure 2 shows how coefficients of autocorrelation cepstral are produced in formula (25). After Fourier transform and logarithm are applied, absolute value is gained and then the result is quadrated. In the end, the reverse Fourier transform is applied.

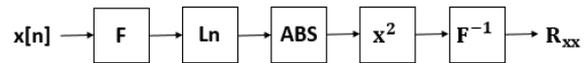

Fig. 2. Displaying the system of cepstral coefficients of correlation suggested for the extraction of echo hiding.

### 5.2. Proposed Security Improvement for Echo Hiding
The pattern can be applied as a general one for other cryptographic methods. When two or more methods are used for steganogarphy, invaders suspect that more methods are used. So, if they find out how to pass the first steganography method, it is complicated to find the second or the order of methods. It is also difficult for invaders to find the number of methods. If the invaders use temporary uncovering, reaching messages will be even more complicated.

In the pattern suggested for increasing the security of echo hiding, we found a method more secure than echo hiding [31]. This method is based upon the key. Therefore, a method is first defined to produce a series of keys, and then we discuss the integration method.

### 5.2.1. How to Produce Sub-keys
An important characteristic for producing a series of keys is the length. A key should be long enough and include non-repetitive keys. In producing the proposed key method, the sender and the receiver should first agree on the primary key; then a matrix with fixed numbers, whose rows and columns are equal to the columns of the primary key, is provided for both parties of steganogarphy.

Therefore, in order to produce subkeys, we act as the following (algorithm):

1. First, the primary key is multiplied in constant matrix; it is a decimal multiplication.

2. Then, the result is multiplied in binary number; the result of this process is the first sub key (the ith).

3. We then shift both the primary key and the constant matrix rows to the right.

4. To produce the next keys, we repeat stages 1 to 3 until the total number of bytes equals the number of bytes of the hidden message.

It should be mentioned that primary key and constant matrix should not be necessarily binary. Primary key only needs to be a series of numbers, and displacement to the right should be equal to numeral displacement, which is shown in the following example:

Example 1) If the primary key is according to string 2531 and constant matrix in the first line 1260, the second line 3001, the third line 1021, and the last line 2021, the shifted key will equal to 1253 and shift to the right for each line in matrix is 0126, 1300, 1102, and 1202, respectively. In the process of multiplication, at first 2531 is multiplied by 0111, which is the first column, then the same number is multiplied by 6022 in the second column. We continue up to the last column, and finally we sum up the results. The sum is equal to 96839327, whose binary equivalent is 101 1100 0101 1010 0110 1001 1111.

Since steganogarphy in VoIP is immediate and the time of execution is very important, key expansion steganogarphy algorithm of advanced encryption standard (AES) cannot be simultaneous with steganography. The only condition for it is a sufficient number of keys that have been produced and shared with the sender and the receiver, before steganography.

It is also possible to choose a key which is big enough to seem pseudo-random as the main one, but it is hard to memorize and use such a key.

In a voice communication based on VoIP, speech is divided into packs to be transferred on th Internet network. Each pack contains a few samples of speech as the main data, and a head containing a series of routing information [32]. Therefore, in the suggested steganography on VoIP, we need sufficient long sub keys so that we can cover all packs and lengthy hidden messages.

In Figure 3, the procedure of key production is shown. In i=0 the primary key is multiplied by the constant matrix, then the result is placed as the variable of sub-keys. With any increase in "i" value in every stage, the primary key and constant matrix shift to the right by primary-key-shift and the constant-matrix-shift functions, respectively. Next, the primary key is multiplied by constant matrix again, and the result is transferred to the sub key variable. The circle continues till "i" variable is smaller than or equal to the length of hidden message. At every stage, the result of sub-key is changed from decimal to binary number (binary Conversion) for transfer.

Mathematical presentation of the exact procedure of sub-key production is:

$$sub_{key_i} = \begin{cases} primary_{key} * constant_{matrix} & i = 0 \\ primary_{key}^{K_i} * constant_{matrix}^{K_i} & i != 0 \end{cases} \quad (26)$$

Where "i" shows every shift from the primary-key to the constant-matrix.

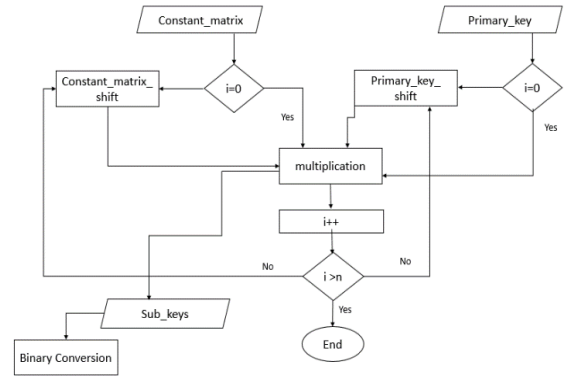

Fig. 3. The procedure of sub-key production using primary-key.

## 5. 2. 2 Proposed embeded method:

First. the hidden message was encrypted. The process of encryption can happen at the same time with steganogarphy using linear feedback shift register (LFSR). LFSR is considered as a strong encoder which produces pseudo-random numbers, in some cases with great length. But in order to improve performance, preferrably first the hidden message should be encoded. After the sub-keys are produced in binary form to be embeded, we should check the binary string of first sub-keys (that is the first sub-key produced using the algorithm as shown in Fugure3). If the first byte of the first sub-keys equals 1, then we should apply the algorithm of spread spectrum for the first frame of sound. And if the byte is equal to 0, the algorithm of echo hiding is used for the first frame of sound. Then, the second byte of the first sub-key and the second frame of the sound are checked. We continue the process for the next bytes until the hidden message or the first sub-key ends. If the hidden message ends before the first sub-key, the number of the sub-key is sent to the receiver using the first integration method (spread spectrum). Otherwise, the first sub-key is finished and we use the second key. The process continues until all bytes of hidden message are finished.

In Figure 4, the process of integration is shown. In this figure, we checked the bytes of the sub-key considering its value. The frame of speech is encrypted with one of the methods. At the same time, we checked the embedding process of the completed hidden messages. For embedding action, we have the following:

$$y(i) = \begin{cases} x(i) + Abs & sub_{key_i(j)} = 0 \\ x(i) + \alpha x(i-d) & sub_{key_i(j)} = 1 \end{cases} \quad (27)$$

In which "i" is the index of frame and j shows the number of byte from the sub-key "i". x(i) Is a frame of speech on which steganography is applied through echo hiding or spread spectrum, and y (i) is steganographic.

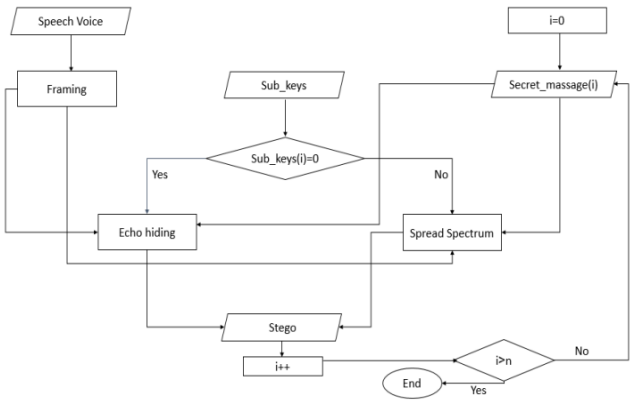

Fig. 4. The embedding process according to proposed method.

### 5.2.3. Proposed Extraction Method

At the extraction stage, we will do what we did at the integration stage. First, the receiver who is aware of steganography, produces a sub-key, because the primary key and the algorithm of sub-key have been shared with the sender and the receiver. Since every byte of sub-key is 0 or 1, the first or second extraction algorithm is used to extract the hidden message. We continue this process until we get the number of byte and sub-key which were also hidden using the first algorithm.

Figure 5 shows the process of extraction. Here, the byte of the sub-key determines which algorithm should be used. The mathematical relationship between these elements is as follows:

$$x(i) = \begin{cases} sgn\left(y^{T}s\right) & sub_{key,(j)} = 0 \\ FT^{-1}\left[\left|\ln\left(FT\left[\,y(n)\right]\right)\right|^{2}\right] & sub_{key,(j)} = 1 \end{cases} \qquad (28)$$

In this equation, "i" is the index of frame, "j" shows the number of byte from sub key i, x (i) is hidden message and y (i) is an encrypted frame of speech. The first part is the extraction of spread-spectrum method, upon whose occurence subkey equals zero. In the second part, the proposed extraction for the echo hiding method occurs when the byte is 1.

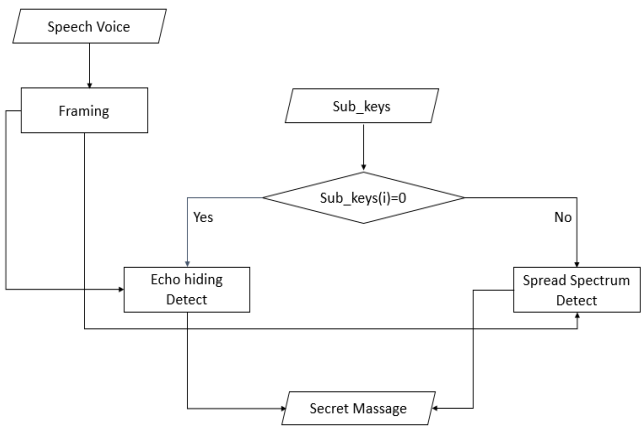

Fig. 5. The Extraction procedure according to the proposed method.

### 6. Simulation and Results

After the proposed method is explained, in this section we will try to show the results of its improved version through experiments and making comparisons. This section is divided into two parts. Part 6.1 deals with the results and comparisons regarding improvements used on the extraction part of echo hiding. Part 6.2 shows the results and comparisons made about the security of this method.

### 6. 1 Improvement in extraction step

As we mentioned earlier, we first aim to improve the extraction stage for echo hiding. This idea has gone through multiple tests after assimilation, which will be discussed later.

#### 6.1.1 The rate of message recovery

For the rate of message recovery, we have the following:

$$Rate = \left(N_{c} \,/\, N\right) \times 100 \quad , \qquad (29)$$

In which $N_{c}$ shows the number of bytes of messages recovered correctly, and N shows the total number of bytes. The unit for the above equation is percentage. Therefore, we can calculate the rate of messages recovery for two different methods of extraction despite different echoes; the results in Table 1 show the rate of message recovery in echo hiding method and also Table 2 shows the rate of message recovery according to the proposed method for series of data with different values of ɑ.

In these tables, for both methods duration of delay for bytes of 0 and 1, the number of bytes for hidden messages and also the framing of speech are considered the same. A set of ten audio data from data center TIMIT in [33] has been brought, each of which has 6000000 samples and have been tested with .mat format (equal to 375 seconds with sampling rate of 16/kHz). The length of message is equal to 10000 bytes and the length of each frame is 600 (equivalent to 0.0375 second, sampling rate of 1600byte/s). To produce the bytes of message, a function, which produces random numbers, has been used. It is obvious that the rate of message recovery improves by increasing the amount of echo in both methods, but the amount of security decreases.

Table 1. The Rate of Message Recovery in Echo Hiding Method

| α Data | 0.05 | 0.1 | 0.2 | 0.3 | 0.4 | 0.5 |
|---|---|---|---|---|---|---|
| D1 | 60.3 | 67.9 | 81.5 | 88.5 | 92.8 | 95.1 |
| D2 | 59.2 | 68.1 | 80.7 | 88.1 | 92.8 | 95 |
| D3 | 59.7 | 67.9 | 80.9 | 88.2 | 92.2 | 95 |
| D4 | 59.5 | 67.7 | 79.9 | 87.1 | 91.5 | 94.2 |
| D5 | 58.8 | 67.4 | 80.4 | 88.5 | 92.7 | 95 |
| D6 | 59.4 | 67.6 | 80.9 | 88.3 | 92.2 | 95 |
| D7 | 59.2 | 68.3 | 80.6 | 87.9 | 92.5 | 94.7 |
| D8 | 58.6 | 68.9 | 80.2 | 87.8 | 91.8 | 94.5 |
| D9 | 60.1 | 67.7 | 80.2 | 88.1 | 93.2 | 95.4 |
| D10 | 58.7 | 66.6 | 78.3 | 86.7 | 92.1 | 94.5 |
| Average | 59.34 | 67.81 | 80.4 | 87.91 | 92.4 | 94.83 |

Table 2. The Rate of Message Recovery Based on Proposed Method

| A\Data | 0.05 | 0.1 | 0.2 | 0.3 | 0.4 | 0.5 |
|---|---|---|---|---|---|---|
| D1 | 60.3 | 67.9 | 81.5 | 88.5 | 92.8 | 95.1 |
| D2 | 59.2 | 68.1 | 80.7 | 88.1 | 92.8 | 95 |
| D3 | 59.7 | 67.9 | 80.9 | 88.2 | 92.2 | 95 |
| D4 | 59.5 | 67.7 | 79.9 | 87.1 | 91.5 | 94.2 |
| D5 | 58.8 | 67.4 | 80.4 | 88.5 | 92.7 | 95 |
| D6 | 59.4 | 67.6 | 80.9 | 88.3 | 92.2 | 95 |
| D7 | 59.2 | 68.3 | 80.6 | 87.9 | 92.5 | 94.7 |
| D8 | 58.6 | 68.9 | 80.2 | 87.8 | 91.8 | 94.5 |
| D9 | 60.1 | 67.7 | 80.2 | 88.1 | 93.2 | 95.4 |
| D10 | 58.7 | 66.6 | 78.3 | 86.7 | 92.1 | 94.5 |
| Average | 59.34 | 67.81 | 80.4 | 87.91 | 92.4 | 94.83 |

For better comparison, the results of Tables 1 and 2 are averaged and shown in Figure 6. In this figure, continuous the line shows the proposed method and the dots show echo hiding method [3].

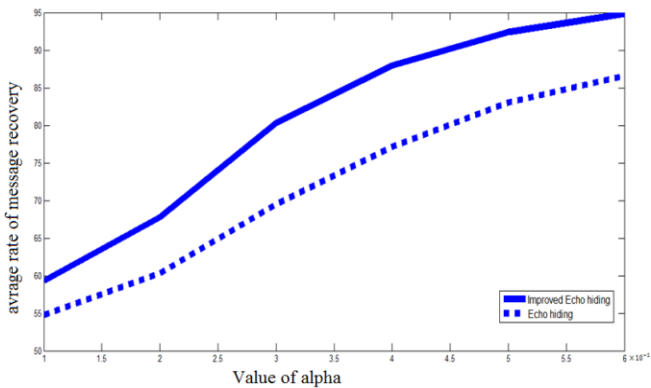

Fig. 6. The comparison made on both methods for the average rate of message recovery

Regarding the given amount in Tables 1 and 2, it is observed that the closer α gets to 1, the more the rate of message recovery increases and inclines toward 100.

It has also been proved that when the amount of echo is less than 0.31, it is intangible. If its domain is not higher than 0.47, the embedded echo is tangible but not annoying [33]. Therefore, the rest of attacks occurs when α=0.3

### 6.1.2. The rate of message recovery in case of Gaussian noise attack

The signal is usually influenced by noise on the way to be sent. In order to evaluate the signal strength against the noise, white Gaussian noise is added to the steganographic signal, considering the nature of ordinary noise. In this test, first white Gaussian noise is added, and the ratio of signal to the noise is specified between 25dB to 40Db. Next, random normal numbers with an average of 0 and variance of 1 are added to the sample signal. Table 3 shows that the rate of message recovery in both methods in case of white Gaussian noise attack for echo is 0.3

Table 3. The Rate of Message Recovery in Case of White Gaussian Attack for α=0.3

| Data | Echo hiding method [3] | The proposed method |
|---|---|---|
| D1 | 77.98 | 88.5 |
| D2 | 77.5 | 87.97 |
| D3 | 76.60 | 88.16 |
| D4 | 77.04 | 87.17 |
| D5 | 76.93 | 88.09 |
| D6 | 77.28 | 88.46 |
| D7 | 76.70 | 87.85 |
| D8 | 77.04 | 87.89 |
| D9 | 76.78 | 88.36 |
| D10 | 76.52 | 86.45 |

### 6.1.3. The rate of message recovery in case of compression attack

In steganogarphy methods based on sound, the ability to cope with compression is tested using MP3 compression. MP3 is an acronym for the third layer MPEG-1 and belongs to compression standards of MPEG-1 from the family of voice coding MPEG. It is a lossy method that eliminates inconceivable parts from the audio file according to the characteristics of human audio system. In compression test, first the encrypted audio file is compressed with a rate of mkbps by the encoder. The amount of m can be 64, 96,128. Therefore, if main steganographic signal, whose format is mostly wav, is attacked, it is first turned into MP3 format, and then MP3 decoder removes its compressed form and again it is turned into wav format. This process destroys a part of sound. The results of message recovery rate are shown in Table 4 for the data gathered through both echo hiding and extraction methods.

Table 4. The Rate of Message Recovery in Case of Compression Attack α=0.3

| Data | Steganography method [3] | The proposed method |
|---|---|---|
| D1 | 63.47 | 73.89 |
| D2 | 63.24 | 73.31 |
| D3 | 61.83 | 73.03 |
| D4 | 61.37 | 72.92 |
| D5 | 62.19 | 72.27 |
| D6 | 62.89 | 73.82 |
| D7 | 62.01 | 72.17 |
| D8 | 62.41 | 71.74 |
| D9 | 61.20 | 72.54 |
| D10 | 62.27 | 71.71 |

### 6.1.4. The rate of message recovery in case of low pass filtering attack

Low pass filter can significantly eliminate the message. Signal disconnection in low pass filters is more considerable than that of high pass filter, because low frequency components of the audio signal are more perceptible. In other words, low frequency of talk signal comprises a sound with low tone and it is impalpable. High frequency of talk signal comprises a sound with high tone. In a conversation people almost talk with a normal tone and sometime they talk loudly. For example, the frequency range of human speech is between 100Hz to 8 KHz; this range mostly leans towards high frequency; therefore, parts with high frequency can be filtered more than parts with low frequencies. This means that only low frequencies can pass and high frequencies can be eliminated. On the other hand, echo hiding works better in high frequencies; therefore, low pass filter attack leads to heavy destruction of this method. The rate results of both methods in case of low pass filter attack is 0.3 and shown in Table 5.

Table 5. The Rate of Message Recovery in Case of
Low Pass Filter Attack for α=0.3

| Data | Echo hiding [3] | The proposed method |
|------|-----------------|---------------------|
| D1 | 63.67 | 70.44 |
| D2 | 63.88 | 70.89 |
| D3 | 62.67 | 70.59 |
| D4 | 63.87 | 70.33 |
| D5 | 63.26 | 69.93 |
| D6 | 63.44 | 71.28 |
| D7 | 62.64 | 70.50 |
| D8 | 62.80 | 69.53 |
| D9 | 63.35 | 71.51 |
| D10 | 63.04 | 70.89 |

### 6.1.5. The rate of message recovery in case of resampling attack

The rate of sampling Fs is a parameter in digital signals. Therefore, the frequency of sampling signal, which is steganographic, is also Fs. To do resistance test against further sampling, the rate of sampling for steganographic signal is first decreased to Fs/2 or Fs4 and then turned back to the same Fs. The rate results of message recovery for both methods in case of attack is 0.3 as shown in Table 6.

Table 6. The Rate Result of Message Recovery in Case
of Further Sampling Attack for α=0.3

| Data | Echo hiding [3] | The proposed method |
|------|-----------------|---------------------|
| D1 | 63.78 | 75.30 |
| D2 | 63.25 | 74.38 |
| D3 | 62.08 | 74.01 |
| D4 | 61.69 | 73.69 |
| D5 | 62.42 | 73.61 |
| D6 | 63.09 | 74.90 |
| D7 | 62.60 | 73.32 |
| D8 | 62.09 | 73.19 |
| D9 | 61.99 | 74.09 |
| D10 | 62.06 | 72.34 |

### 6.1.6. The rate of message recovery in case of Requantization attack

In steganographic voice, the level of signal samples has been chosen carefully with 16 bytes. In this attack, the accuracy for sampling level has been reduced to 8 bytes for each sam-ple and then turned back to 16 bytes. The rate results of message recovery in both methods in case of accuracy reduction attack for echo is 0.3 as shown in Table 7.

Table 7. The Rate of Message Recovery in Case
of Accuracy Reduction Attack for α=0.3

| Data | Echo hiding [3] | The proposed method |
|------|-----------------|---------------------|
| D1 | 63.80 | 75.26 |
| D2 | 63.27 | 74.30 |
| D3 | 62.3 | 74.05 |
| D4 | 61.81 | 73.66 |
| D5 | 62.42 | 73.63 |
| D6 | 63.05 | 74.85 |
| D7 | 62.53 | 73.31 |
| D8 | 62.07 | 73.19 |
| D9 | 61.98 | 74.08 |
| D10 | 62.05 | 72.30 |

### 6.2. Improvement the security step

The second idea in this article is to improve the security of echo hiding method, which will be discussed later after the results of algorithm assimilation are presented. Before discussing assimilation in this section, we will first check the capacity of the proposed method. Spread spectrum algorithm and echo hiding both have only one steganographic byte in each frame; therefore, in each frame only one byte is steganographic. On the other hand, the capacity of these methods in VoIP, which is equal to 20byte/s; thus, the capacity of the proposed method in VoIP is also 20byte/s.

The simulation of the sections from 6-2-1 to 6-2-5 has been carried out under similar conditions. Message, primary key, and constant matrix have been produced by functions generating pseudo-random numbers. The length of hidden message is 10000 bytes. The primary key has 100 numbers and the constant matrix is also a 100*100 matrix. The key for spread spectrum for each execution is formed by a function generating pseudo-random numbers. In addition, the average time complexity of each run is 1.2 seconds in 59 runs.

In the method proposed, if the number of speech samples on the length of the hidden message is less than a threshold value, steganography for VoIP does not happen. This threshold value is the same limit, which defines the border between loyalty to the loud sound and capacity. If the limit is too small, the capacity is increased, whereas if it is too big, it leads to increase in loyalty to the loud sound and security increment. In this assimilation, the length of each frame is considered higher than 20 samples on second.

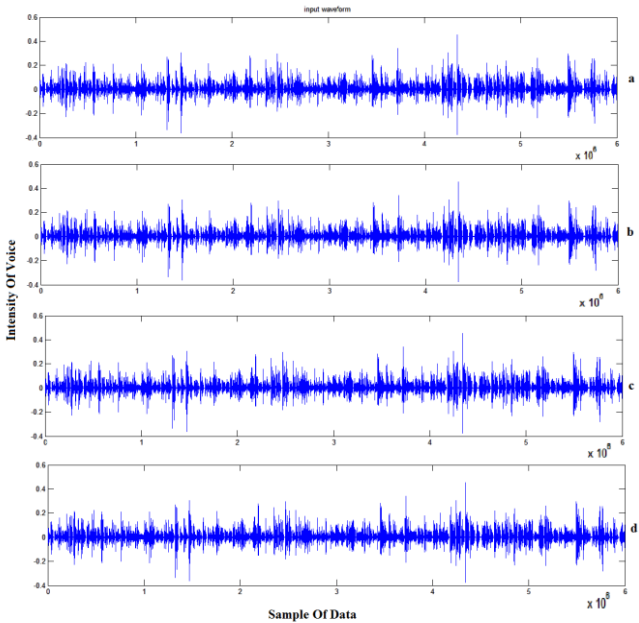

Fig. 7. Sound samples: primary sound, suggested steganography, echo, and spread spectrum

The outputs gained through these sound samples are primary sound, suggesting steganography, echo hiding, and spread spectrum, respectively (see Figure 7).

### 6.2.1 The rate of message recovery
Rate of message recovery for the proposed method can be gained using equation (24) in order to improve the security of the method. The numbers presented in Table 8 shows approximately the rate of message recovery, which is different by 2% in each performance. This value is gained through a series of data for echo hiding, spread spectrum, and the proposed method.

Table 8. The Rate of Message Recovery

| Data | Echo hiding [3] | Spread spectrum [24] | The proposed method |
|------|-----------------|----------------------|---------------------|
| D1 | 80.67 | 89.82 | 88.9 |
| D2 | 80.69 | 93.89 | 89.09 |
| D3 | 80.59 | 94.36 | 89.06 |
| D4 | 79.84 | 93.72 | 88.74 |
| D5 | 80.42 | 94.43 | 89.1 |
| D6 | 80.52 | 93.89 | 88.91 |
| D7 | 80.2 | 93.95 | 88.54 |
| D8 | 80.2 | 93.66 | 88.22 |
| D9 | 80.24 | 94.08 | 89.55 |
| D10 | 78.31 | 93.47 | 88.13 |

### 6.2.2. Message recovery in case of white Gaussian noise
Regarding the explanation provided in section 5.1 for checking the strength of the method, we only obtained the rate of message recovery in case of white Gaussian noise, because it is one of the most common attacks in VoIP channels. Hence, only this attack is assimilated to check the strength of the method. Table 9 shows the rate of message recovery in three methods in case of white Gaussian noise attack.

Table 9. The Rate of Message Recovery in Case of white Gaussian Attack

| Data | Echo hiding [3] | Spread spectrum [24] | The proposed method |
|------|-----------------|----------------------|---------------------|
| D1 | 80.59 | 89.74 | 87.34 |
| D2 | 80.6 | 93.86 | 87.89 |
| D3 | 80.79 | 94.3 | 87.54 |
| D4 | 79.81 | 93.64 | 87 |
| D5 | 80.48 | 94.4 | 87.37 |
| D6 | 80.49 | 93.85 | 87.45 |
| D7 | 80.2 | 93.9 | 86.81 |
| D8 | 80.14 | 93.63 | 86.67 |
| D9 | 80.17 | 93.97 | 87.99 |
| D10 | 78.3 | 93.47 | 86.69 |

### 6.2.3. SNR rate (signal to noise ratio)
Another rate reviewed is SNR which is used for the evaluation of system performance. SNR assesses the ratio between the strength of a cover data and a noise power and is obtained by the following formula (30):

$$SNR_{db} = \frac{P_{signal}}{P_{noise}}. \tag{30}$$

SNR gained through spread spectrum, echo hiding, and the proposed method is shown in Table 10. The results of SNR gained through positive and negative echo hiding, backward and forward echo hiding, and the mirrored echo hiding are all shown in Table 11. Table 12 also contains the results of the same tests on methods such as time-spread echo hiding, and improved spread spectrum. In every execution, each of these methods varies up to 0.1.

If the amount of SNR increases, the quality of final voice or speech is higher. In speech, if SNR rate is 20 or above, the quality is good. As can be observed, the method proposed can lead to a good quality.

Table 10. The Amount of SNR

| Data | Spread Spectrum Method [24] | Echo Hiding Method [3] | The proposed Method |
|------|------------------------------|------------------------|---------------------|
| D1 | 23.72 | 14.56 | 21.43 |
| D2 | 23.61 | 14.58 | 20.59 |
| D3 | 22.80 | 14.63 | 19.55 |
| D4 | 24.26 | 14.60 | 21.44 |
| D5 | 26.64 | 14.58 | 22.51 |
| D6 | 23.14 | 14.61 | 19.79 |
| D7 | 23.64 | 14.62 | 20.69 |
| D8 | 23.48 | 14.66 | 20.20 |
| D9 | 24.27 | 14.61 | 21.49 |
| D10 | 22.47 | 14.62 | 19.72 |

Table 11. The Amount of SNR

| Data | NP Echo hiding[8] | BF Echo hiding[6] | M Echo hiding[9] |
|------|------|------|------|
| D1 | 13.41 | 13.4 | 13.85 |
| D2 | 13.35 | 13.48 | 13.82 |
| D3 | 13.28 | 13.48 | 13.81 |
| D4 | 13.41 | 13.53 | 13.98 |
| D5 | 13.54 | 13.41 | 13.99 |
| D6 | 13.34 | 13.46 | 13.85 |
| D7 | 13.45 | 13.44 | 13.86 |
| D8 | 13.30 | 13.43 | 13.72 |
| D9 | 13.39 | 13.51 | 13.94 |
| D10 | 13.60 | 13.43 | 14.11 |

Table 12. The Amount of SNR

| Data | TS Echo hiding[4] | Improved SS[27] |
|------|------|------|
| D1 | 9.93 | 41.28 |
| D2 | 9.87 | 41.63 |
| D3 | 9.73 | 41.23 |
| D4 | 0.93 | 41.80 |
| D5 | 10.03 | 41.30 |
| D6 | 9.74 | 41.35 |
| D7 | 9.71 | 41.37 |
| D8 | 9.81 | 41 |
| D9 | 9.12 | 41.20 |
| D10 | 9.84 | 41.08 |

### 6.2.4. Reviewing cepstral coefficients

In this section, the security of the proposed method is evaluated. According to Chen [5], the most influential attack on hidden data, which is steganographic by echo hiding, is to detect embedded echo and change it. The status of echo is obtained by calculating cepstral coefficients. Intense changes in cepstral coefficients lead to its detection through echo hiding. One way to increase the security of the method is to decrease the amount of echo. As the amount of echo decreases, the process of message recovery decreases significantly, and the security is relatively increased. Another approach proposed here is to apply the proposed method, which is a combination of echo hiding on the basis of key with spread spectrum method. On the other hand, message recovery is increased and shifts to cepstral coefficients are much less than the previous condition.

In Figure 8, a chart of cepstral coefficients for steganographic signals of proposed method with steganographic signals of echo hiding is in one frame. As it is seen, when echo hiding is used, there are many peaks in cepstral coefficients. This difference exists on other sounds and frames with nuance; the following image shows this nuance. The difference between cepstral coefficients leads to revelation of steganography through the existing steganalysers. However, using the proposed method shows a good resistance for echo hiding against the attacks of existing steganalysers due to the smoothing peaks of cepstral coefficients.

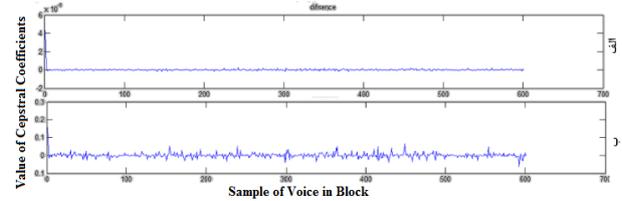

Fig. 8. The cepstral coefficients when (a) the proposed steganography and (b) echo hiding [3] are used.

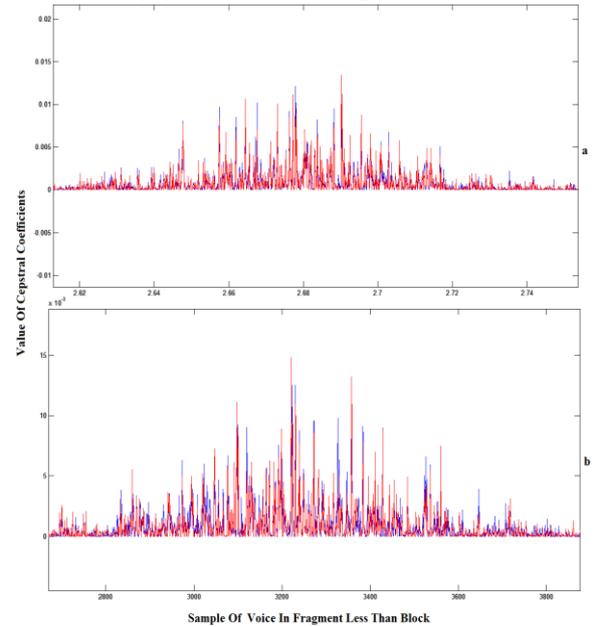

Fig. 9. The cepstral coefficients of covering and steganographic signals when the proposed method (a) and echo hiding (b) [3] are used.

The following images are the samples of cepstral coefficient 2700 to 2800 from audio steganographic signal and covering audio signal. In Figure 9, first the proposed-steganography and then echo hiding are used. The blue diagram shows cepstral coefficient of covering signal, while the red diagram shows cepstral coefficient of steganographic signal. Figures 10, 11, 12 and 13 indicate the cepstral coefficient of covering signal, steganographic signal in positive and negative echo hiding, backward and forward echo hiding, mirrored echo hiding and time-spread echo hiding, respectively. These figures mostly indicate the sample cepstral coefficients from 2500 to 4500. Here, the blue color shows covering signal and the red color shows steganographic signal.

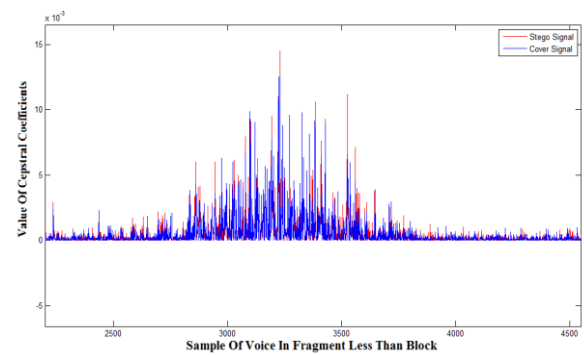

Fig. 10. The cepstral coefficients for covering and steganographic signal when the positive and negative echo hiding [8] is used.

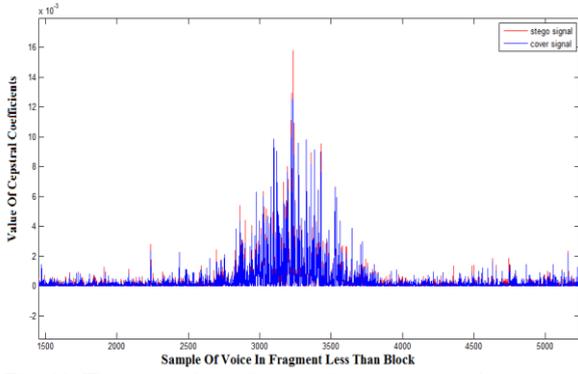

Fig. 11. The cepstral coefficients for covering and steganographic signal when backward and forward echo hiding [6] is used.

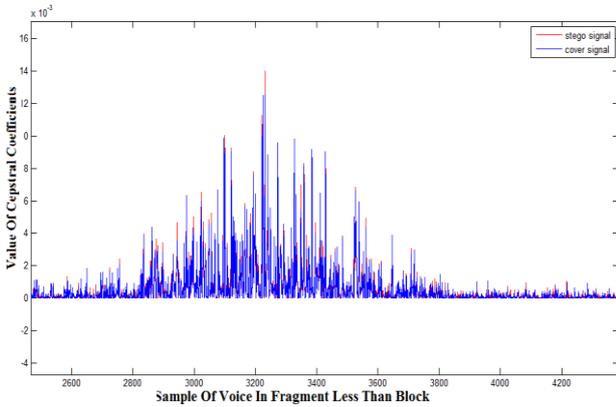

Fig. 12. The cepstral coefficients for covering and steganographic signals when mirrored echo hiding [9] is used.

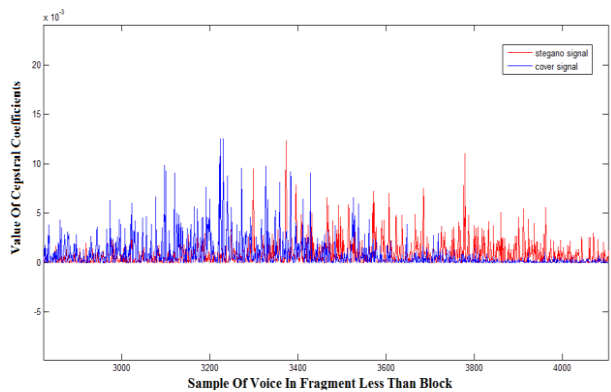

Fig. 13. The cepstral coefficients for covering and steganographic signals when time-spread echo is used [4].

### 6.2.5 Applying MFCC steganalysis to check the security

Mel-cepstrum coefficients or MFCC was first introduced in the 1980s for speech recognition. In 2007, this was used for steganalysis [34]; the following shows how the coefficients were calculated:

$$MFCCs = DFT\left(\log\left(MT\left(DFT(x)\right)\right)\right) = \begin{pmatrix} x_{mel1} \\ x_{mel2} \\ \vdots \\ x_{melC} \end{pmatrix}. \qquad (31)$$

DFT refers to discrete fourier transform and MT frequency maps on the scale of Mel. In the current study, for steganalysis the function of MFCC is used to extract speech characteristics. For categorization and instruction, Gaussian mixture model (GMM) is used. The categorization involves two classes for covering conversation and steganographic conversation with both echo hiding and spread spectrum methods. For instruction, 900 files with 20000 (equal to 1.25 second) samples were considered for each class (that is 8100 files for covering and steganographic classes). For the test, 1000 files and 20000 samples were considered for covering, steganographic and echo hiding (positive or negative), backward and forward steganography, mirrored echo hiding, time-spread, spread spectrum, improved spread spectrum and finally the proposed method. In total, there are 9000 test files. To compare the function of all eight methods, the classification error with equation (32) is used:

$$P_E = \min_{P_{FA}} 1/2\left(P_{FA} + P_{MD}(P_{FA})\right) \qquad (32)$$

In this formula, $P_{FA}$ and $P_{MD}$ are the probability of false alarm and the probability of missed detection, respectively.

Results gained through assimilation were shown in Table (13). In this table, $P_{FA}$ is equal to 0.212 and $P_{MD}$ in echo hiding, spread spectrum and proposed method respectively equal to 0.32, 0.453, and 0.672. $P_{MD}$ in positive and negative echo hiding is equal to 0.24; in backward and forward echo hiding is equal to 0.31; in Mirrored echo hiding is equal to 0.209; in time-spread echo hiding is equal to 0.11 and in improved spread spectrum is equal to 0.47.

Results from the classifier errors indicate that the more classifier error is detected for the method used for steganography, whether it is a steganographic file or not, the safer the method is.

Table 13. The Results of the Classification Error

| Methods | Amount |
|---|---|
| Spread Spectrum [24] | 0.154 |
| Echo hiding [3] | 0.147 |
| NP Echo hiding [8] | 0.131 |
| BF Echo hiding [6] | 0.140 |
| Mirrored echo hiding [9] | 0.128 |
| TS Echo hiding [4] | 0.117 |
| Improved SS [27] | 0.155 |
| Proposed method | 0.177 |

### 7. Conclusion

As was mentioned in Section 5, in this article extraction and security of the echo hiding was improved. Improvement in extraction was carried out by rejecting self-symmetric assumption of the signal. After rejecting the assumption, autocorrelation of cepstral coefficient was again calculated, thus we gained different results. The rate results of the message recovery gained through simulation proved improvements. The improvement in security section was reached by mixing this method with spread spectrum steganography based on the key. In order to generate pseudo-random key generation, a key generation algorithm was also presented in this paper. This algorithm used a primary key as seed. In addition to security improvement, message recovery, signal rate to the noise and the capacity were all improved by this method. To make sure that the security level is upgraded, we compared the proposed method with more recent methods. Results gained through different tests from Sections 6.2.3 to 6.2.5 indicate the superiority of the proposed method over other methods on echo hiding. Other results reached through applying steganalysers show that the proposed method is more secure than the important improved method of spread spectrum. We hope that in future research we can maintain the rate of message recovery as well as its security. The idea to increase the capacity embedded in each frame with both algorithms of steganogarphy is done by one of the key conditions.